\documentclass[12pt]{article}
\usepackage{epsfig}

\usepackage{amsfonts, graphicx,caption,subfig}

\usepackage{amssymb}
\usepackage{amsmath}
\usepackage{amsfonts}

\usepackage{amsthm}

\theoremstyle{theorem}

\theoremstyle{definition}

\def\bp{\begin{proof}}
\def\ep{\end{proof}}

\def\be{\begin{equation}}
\def\ee{\end{equation}}
\def\ba{\begin{array}{c}}
\def\ea{\end{array}}

\def\ben{\[}
\def\een{\]}

\newcommand{\bea}{\begin{eqnarray}}
\newcommand{\eea}{\end{eqnarray}}

\newcommand{\bbr}{\br\!\br}
\newcommand{\kkt}{\kt\!\kt}

\newcommand{\kt}{\rangle}
\newcommand{\br}{\langle}

\begin{document}

\titlepage
.

 \vspace{4.35cm}


 {\Large \bf

Coulomb potential

and the paradoxes of PT-symmetrization\footnote{paper presented
during the Peter Leach's birthday international conference \\
\mbox{\ \ \ \ }  {\bf ``Tercentenary of the
Laplace-Runge-Lenz vector"};\\
\mbox{\ \ \ \ }  Salt Rock Hotel, Durban, November 23-27, 2011.\\}

  }


\vspace{10mm}

 {\bf Miloslav Znojil}

 {Nuclear Physics Institute ASCR,
 250 68 \v{R}e\v{z}, Czech Republic,

 e-mail:
 znojil@ujf.cas.cz}

 \vspace{3mm}

\vspace{3mm}

 \section*{Abstract}

Besides the standard quantum version of the Coulomb/Kepler problem,
an alternative quantum model with not too dissimilar
phenomenological (i.e., spectral and scattering) as well as
mathematical (i.e., exact-solvability) properties may be formulated
and solved. Several aspects of this model are described. The paper
is made self-contained by explaining the underlying innovative
quantization strategy which assigns an entirely new role to
symmetries.

 \begin{center}

\end{center}

\vspace{5mm}

\newpage

\section{Introduction}

The list of the traditional roles of symmetries in Quantum Theory
has recently been enriched by the increasingly active studies of the
role of the so called ${\cal PT}-$symmetries \cite{Hook}. Typically,
one considers a Hamiltonian composed of a kinetic- and
potential-energy terms,
 $$
 H= -\frac{\ \hbar^2}{2m} \,\triangle+V(x)
 $$
and replaces the current requirement $H = H^\dagger$ of its
Hermiticity in Hilbert space $L^2(\mathbb{R}^d):={\cal
H}^{(friendly)}$ by the less trivial physical Hermiticity (or rather
``crypto-Hermiticity'' \cite{SIGMA}) $H = H^\ddagger$ which must be
constructively, {\em ad hoc} defined in {\em another} Hilbert space
${\cal H}^{(standard)}$. In the language of symmetries this means
that, firstly, the ``usual'' Hermiticity  $H = H^\dagger$ is
reinterpreted as a formal time-reversal symmetry
 \be
 H \,{\cal T} = {\cal T}\,H
 \label{hermi}
 \ee
(with a suitable time-reversal operator ${\cal T}$). In the second
step one rejects the constraint (\ref{hermi}) as ``too formal''
\cite{Carl} and postulates, instead, the Hermiticity of $H=H^*$ in
an auxiliary Krein space with a suitable indefinite pseudometric
${\cal P}$. In other words,  in the language of symmetries one
follows the recommendations of mathematicians \cite{BG,Langer} and
re-facilitates the mathematics by the replacement of Eq.~
(\ref{hermi}) by the modified requirement called ${\cal PT}$
symmetry,
 \be
 {\cal
 PT}\,H=H\,{\cal PT}\,.
 \label{ptsh}
 \ee
In the context of physics the recipe finds its most ambitious {\em
theoretical} encouragement in the appeal of the concept of such a
form of symmetry in relativistic quantum field theory \cite{Str},
with ${\cal P}$ representing the parity and with ${\cal T}$
mimicking the time reversal in this implementation.

Another explanation of the increasing popularity of the whole
concept certainly lies in the amazing productivity of the ${\cal
PT}-$symmetrizations of various quantum systems. In the
popularization of such a trick the key role has been played by the
serendipitious letter \cite{BB} in which Bender and Boettcher
proposed and demonstrated that the ${\cal PT}-$symmetrization of a
given local Schr\"{o}dinger equation in single dimension (i.e., a
transition from constraint (\ref{hermi}) to constraint (\ref{ptsh})
at $d=1$) may represent an efficient theoretical tool and way
towards finding new phenomenologically useful Hamiltonians with real
spectra. They also recommended to achieve this simply by the
replacement of the original Hermitian potential $V(x)$ by its
non-Hermitian alternative $V({\rm i}x)$ (thorough review offered by
paper \cite{Carl} may be recommended as an introductory reading).

In our present paper we intend to return to the older application of
this idea to the exactly solvable Coulomb/Kepler problem
\cite{ptcoul} and to its recent upgrades \cite{geza} --
\cite{annals}. We intend to review the related recent theoretical
developments and to show how the incessant progress in the field
applies to this particular but important example.

\section{The ${\cal
PT}-$symmetric version of the Coulomb problem}

In {\em loc. cit.}, the standard quantum Coulomb/Kepler problem has
been assigned a new, non-equivalent quantum system version which is
formally represented by the  ${\cal PT}-$symmetric Schr\"{o}dinger
equation
 \be
 \frac{\ \hbar^2}{2m} \,
  \left [-\frac{{\rm d}^2}{{\rm d}{x}^2}
  +  \frac{L(L+1)}{{x}^2}\right ]\, \Psi({x})
 +\frac{{\rm i}Z}{x} \, \Psi({x}) = E \,\Psi({x})\,,
 \ \ \ L>-\frac{1}{2}\,,
 \ \ \ \ 2L\ \notin \mathbb{Z}\,.
 \label{CoSEor}
  \ee
We shall abbreviate here $2mE/\hbar^2=-k^2$ and emphasize that even
in the presence of the imaginary unit, this equation remains
solvable in terms of the well known confluent hypergeometric
functions,
 \be
 \Psi_1(x)=C_1 \Psi_1(x) +C_2 \Psi_2(x)\,,
  \label{gsolge}
 \ee
 \be
 \Psi_1(x)=e^{-kx}x^{L+1}
 {_1}{F}{_1}(1+L+{{\rm i}Z}/{(2k)},2L+2,2kx)\,,
  \label{gsol1}
 \ee
 \be
 \Psi_2(x)=e^{-kx}x^{-L}
 {_1}{F}{_1}(-L+{{\rm i}Z}/{(2k)},-2L,2kx)\,.
  \label{gsol2}
 \ee
The well known Coulomb -- harmonic oscillator correspondence has
been studied in \cite{ptcoul}. In the role of a formal postulate it
helped us to fix the physical asymptotic boundary conditions which
would remain, otherwise, ambiguous (Ref.~\cite{tobog} and previous
{\it loci citati} should be consulted for all details).

It has been shown \cite{cubic} that all of the models of the class
sampled by Eqs.~(\ref{CoSEor}) + (\ref{gsolge}) may be perceived,
under certain conditions which we explained in Ref.~\cite{SIGMA}, as
fully compatible with the standard postulates of Quantum Mechanics.

Our present paper will expose Eq.~(\ref{CoSEor}) as a special case
of the broader class of Schr\"{o}dinger equations which all
exemplify an extension of quantum model-building strategies. In
Ref.~\cite{SIGMA} we called this approach a ``crypto-Hermitian'' or
``three-Hilbert-space'' quantum mechanics. We shall emphasize here
that the transition from the Hermitian to ${\cal PT}-$symmetric
language is extremely productive while, at the same time, its
multistep nature often leads to conceptual misunderstandings. While
``teaching by example'', we shall try to clarify here some of the
most blatant ones.

\section{The abstract formalism}

Several compact review papers \cite{Carl,SIGMA,ali} may be
recommended for reference. At the same time, an introductory
explanation of the structure of Quantum Mechanics using the
pseudometric in Krein space may be given, for our present purposes,
a much shorter form.

First of all the readers should be warned that in the majority of
textbooks on Quantum Mechanics the meaning of the Dirac-ket symbols
$|\psi\kt \in {\cal V}$ is merely explained via examples. Most often
one deals just with the most common quantum motion of a point
particle inside a local potential well. Thus, it is assumed that
there exists an operator $\hat{Q}$ of the particle position with
eigenvalues $q\in \mathbb{R}^d$ and eigenkets $|q\kt$. The standard
(often called Dirac's) Hermitian conjugation is then represented by
the transposition plus complex conjugation of any ket vector,
yielding the bra vector,
 \be
 {\cal T}^{(Dirac)}: |\psi\kt \ \to \ \br \psi|\,.
 \label{diraco}
 \ee
In the basis  $|q\kt$ one may then represent the ket of any
pre-prepared state $|\psi\kt$ by the overlap $\psi(q) = \br
q|\psi\kt$, i.e., by the square-integrable wave function of the
coordinate $q$.

Our present purpose is not a criticism of this approach as such
(interested readers may find such a criticism elsewhere \cite{Bohm})
but rather just of one of its consequences. In virtually all of the
similar classes of examples, indeed, the vector space of states
${\cal V}$ is simply assumed endowed with the most common inner
product
 \be
 \br \psi_a|\psi_b\kt = \int_{-\infty}^\infty\,\psi_a^*(q)\psi_b(q)\
 dq
 \label{calling}
 \ee
with, possibly, the integration replaced by the infinite or finite
summation. Thus, we may (and usually do) set ${\cal V}\equiv
L^2(\mathbb{R})$, etc.

Without any real danger of misunderstanding we may speak here about
the ``friendly" Hilbert space of states ${\cal H}^{(F)} \equiv {\cal
V}$, calling the variable $q$ in Eq.~(\ref{calling}) ``the
coordinate''. In parallel we usually perform a maximally convenient
choice of the Hamiltonian $H$ based on the so called principle of
correspondence which encourages us to split the Hamiltonian into the
kinetic and potential energies, $H=T+V$. Whenever the general
interaction operator $V$ is represented, say, by a kernel $V(q,q')$
when acting upon the wave functions, this kernel is most often
chosen as proportional to the Dirac's delta-function so that $V$
becomes an elementary multiplicative operator $V =V_{local}= V(q)$.
Similarly, the most popular and preferred form of the ``kinetic
energy" $T$ is a differential operator, say, $T = T_{local}=
-d^2/dq^2$ in single dimension and in the suitable units.

The word of strong warning emerges when we perform a Fourier
transformation in ${\cal H}^{(F)} $ so that the variable $q$ becomes
replaced by $p$ (= momentum). One should rather denote the latter,
Fourier-image space by the slightly different symbol ${\cal H}^{(P)}
$, therefore (with the superscript still abbreviating ``physical"
\cite{SIGMA}).

Paradoxically, after the latter change of frame the kinetic operator
$T_{local}$ becomes multiplicative while $V_{local}$ becomes
strongly non-local in momenta. Nevertheless, all this does not
modify the overall paradigm. A truly deep change of the paradigm
only comes with the models where the necessity of the observability
of the coordinate $q$ is abandoned completely. One may still start
from the vector space of kets ${\cal V}$ but it makes sense to endow
it with {\em another} Hilbert-space structure,  via the inner
product defined by an integral over a complex path,
 \be
 \br \psi_a|\psi_b\kt = \int_{\cal C}\,\psi_a^*(s)\psi_b(s)\
 ds\,.
 \label{recalling}
 \ee
This is one of the most characteristic intermediate steps made in
the so called ${\cal PT}-$symmetric quantum theories \cite{Carl}.
The resulting loss of simplicity of the position operator $\hat{Q}$
changes the physics of course. The key point is that we lose the
one-to-one correspondence between the integration path ${\cal C}$
and the spectrum $\mathbb{R}$ of any coordinate-mimicking operator.
The physics-independent optional variable $s$ becomes purely formal.

In such a setting our choice of the physical observables must still
obey the old quantization paradigm, which is just set in a modified
context. The loss of the observability of the coordinate proves
essential, anyhow. For illustration one might recall the
pedagogically motivated paper~\cite{Hoo} in which, in a slightly
provocative demonstration of the abstract nature of quantum theory,
the variable $s$ in Eq.~(\ref{recalling}) has been interpreted as an
observable ``time" of a hypothetical ``quantum clock" system.

Once we wish to understand our Coulombic Schr\"{o}dinger eigenvalue
problem (\ref{CoSEor}), we must make one more step and generalize
further the inner products (\ref{recalling}). Such a second-step
generalization of the inner product will certainly move us from the
two Hilbert spaces ${\cal H}^{(F)}$ and ${\cal H}^{(P)}$ to the
third one, viz., to the final and physical ``standard'' Hilbert
space ${\cal H}^{(S)}$ (this notation is taken from
Ref.~\cite{SIGMA}).

The introduction of the third Hilbert space forms the theoretical
background of an amendment of the traditional quantum mechanics, the
key nonstandard features of which can be seen

\begin{itemize}

\item
in the admissibility of the complex potentials sampled by the
power-law-anharmonic family $V(s) = -({\rm i} s)^{2+\delta}$ of
Ref.~\cite{BB} and generating the real and discrete bound-state
spectra at any $\delta>0$ (cf. the proofs in \cite{DDT});

\item
in the replacement of the usual real line of $s$ by a complex curve
${\cal C}=\mathbb{C}(\mathbb{R})$ which may even be, in principle,
living on a complicated multisheeted Riemann surface \cite{tobog};

\item
in the theoretical imperative of the construction of certain
operator $\Theta$ (see below);

\item
in the possibility of a systematic study of the discretizations and
simplifications.

\end{itemize}

 \noindent
In our present paper, the emphasis will be put on the last feature.

At the stage of development where we did not yet explain the meaning
and role of the operator $\Theta$ (called Hilbert space metric) the
theory remains incomplete. We already cannot rely upon a more or
less safe guidance of quantization as offered by the principle of
correspondence. Just a partial revitalization of such guidance is
possible in the new context (cf., e.g., a nice example-based
discussion of this point in Ref.~\cite{cubic}).

This being said, the main theoretical obstacle lies in the vast
ambiguity of the necessary appropriate generalization of the
Hermitian conjugation as prescribed by Eq.~(\ref{diraco}). The
general recipe (explained already in \cite{Geyer} or, more
explicitly, in \cite{SIGMA}) is $\Theta-$dependent and reads
 \be
 {\cal T}^{(general)}_\Theta: |\psi\kt \ \to \ \bbr \psi|:=\br
 \psi|\Theta\,.
 \label{nediraco}
 \ee
This means that using the language of wave functions $\psi(s)$ with
$s \in {\cal C}$ we must replace the most common single-integral
definition (\ref{recalling}) of the inner product in the original
``friendly" Hilbert space ${\cal H}^{(F)}$ by the more sophisticated
double-integral formula
 \be
 \bbr \psi_a|\psi_b\kt = \int_{\cal C}\,\int_{\cal C}\,
 \psi_a^*(s)\,\Theta(s,s')\,\psi_b(s')\
 ds\,ds'\,.
 \label{birecalling}
 \ee
In terms of an integral-operator-kernel representation
$\Theta(s,s')$ of our abstract metric operator
$\Theta=\Theta^\dagger>0$ this recipe defines the inner product
which converts {\em the same} ket-vector space ${\cal V}$ into {\em
the amended} and final, metric-dependent and physics-representing
Hilbert space ${\cal H}^{(S)}$ of the very standard quantum theory
(cf. \cite{SIGMA} for more details).

\section{The upgraded formalism  in applications}

In the attempted applications of all of the new ideas to the
traditional benchmark models like Coulomb scattering one may make
use of its traditional merits (like, e.g., exact solvability) as
well as of the flexibility of the choice of
 the ${\cal PT}-$symmetric (i.e., complex and left-right symmetric)
 integration path, sampled in Ref.~\cite{siegl} as follows,
 \be
 x(s)=x^{U}_{(\varepsilon)}(s)\,= \,
 \left \{
 \begin{array}{ll}
 -{\rm i}(s+\frac{\pi}{2}\varepsilon)
 -\varepsilon, & s \in (-\infty, -\frac{\pi}{2}\varepsilon),\\
 \varepsilon e^{ {\rm i}({s/\varepsilon+3/2\pi }
  )},&  s \in
 ( -\frac{\pi}{2}\varepsilon,
 \frac{\pi}{2}\varepsilon),\\
  {\rm i}(s-\frac{\pi}{2}\varepsilon)+\varepsilon\,, & s \in
(\frac{\pi}{2}\varepsilon, \infty).
 \ea
 \right .
 \label{urva}
 \ee
This leads to new results of course. Typically, in spite of the
non-unitarity of the scattering (remember that the Coulomb potential
is strictly local!) the bound-state energies still emerge from the
poles of the scattering matrix \cite{siegl}.

\subsection{Discretizations}

The use of discretizations of the differential forms of
Schr\"{o}dinger operators may be, typically, Runge-Kutta-inspired.
In practice, they are slowly becoming useful in solid-state physics
\cite{Schomerus}, optics
 \cite{Longhi}
and statistical physics \cite{Scott}. Less expectedly, the use of
lattice models proved crucial for the unitarity  of the scattering.
It has been shown \cite{scatt} that the theory of scattering by
non-Hermitian obstacles may be made unitary and consistent via a
certain selfconsistently prepared transition to non-local
potentials. Unfortunately, it is not yet clear how such a
requirement of selfconsistency  could be realized in the continuous
limit of the discrete models.

Whenever we discretize the coordinates and replace the differential
Hamiltonians by matrices with property $H \neq H^\dagger$, the
above-reviewed theory applies without changes. The usual Hilbert
space becomes unphysical and it must be replaced by its unitarily
non-equivalent correct alternative ${\cal H}^{(S)}$ endowed with a
sophisticated metric $\Theta=\Theta^{(S)}\neq I$ which defines the
{\em ad hoc} inner product.

In the discretized version of the theory, the integral kernel of the
metric must merely be replaced by a matrix. Naturally, also the
double integral (\ref{birecalling}) gets replaced by the double sum,
 $$
(\psi,\phi)^{(S)}=\sum_{j,k=1}^N\,\psi^*_j\,\Theta_{j,k}^{(S)}\,\phi_k\
 $$
in
 ${\cal H}^{(S)}$. In this setting,
the imaginary choice of the Coulomb coupling may still be made
compatible with the standard postulates of Quantum Theory, provided
only that it still generates the real, i.e., potentially observable
spectrum of the bound-state energies.

Once we started our considerations from the imaginary Coulomb model
defined along a continuous complex trajectory, we may expect that
many of its properties will survive also the transition to its
discrete descendants. For inspiration
 we may recall Ref.~\cite{siegl} where the bound-state
energies were shown to coincide with the poles of transmission
coefficients. Still, as long as the potential $V \neq V^\dagger$ is
local, the
 {unitarity of the scattering}
 cannot be required \cite{siegl,scatt}.
At the same time, the unitarity of the time evolution of the system
itself may be achieved. Indeed, although the Hamiltonian $H$ is
non-Hermitian in ${\cal H}^{(F)}$, (abbreviated
 $H\neq H^\dagger$), it is Hermitian
 in
 ${\cal H}^{(S)}$.
 This feature is called {cryptohermiticity}, requiring $H=H^\ddagger$ {\em alias}
           $$ H^\dagger \Theta^{(S)}= \Theta^{(S)}\,H\,.$$
Here, the operator or matrix $\Theta^{(S)}$ is precisely the one
which defines the physical inner product.

This being said, the loss of easy constructions {\em is} a problem
\cite{Jones}. Still, the {discretization} of the coordinates may be
recommended as {\em the} recipe.

\subsection{Interpretations}

The {generic  ${\cal PT}-$ symmetric quantum model} describes a
closed {system} {defined} {via a doublet} of operators $ \
{H,\Theta}\ $
  or via a triplet of operators (adding a new observable $\Lambda $
and having, typically, Hamiltonian $H \neq
 H^\dagger$ accompanied by a charge \cite{Carl}), etc.
In other words, the {\em dynamical} content of phenomenological
quantum models is encoded {in} Hamiltonian $H$ {and in} metric
$\Theta$. In this setting the metric $\Theta$ {guarantees the
{unitarity}} of time evolution
 in an
  {\em ad hoc}, ``standard'' Hilbert space \cite{Geyer}, to be denoted by
  the symbol
  ${\cal H}^{(S)}$ in what follows.
In addition, one can also impose some other, phenomenologically
motivated requirements like a {short-range smearing} of coordinates
\cite{fund}, etc.

One of the remarkable features of such an upgrade of applications of
quantum mechanics may be seen in the robust nature of its ``first
principles'' which remain unchanged. Thus, its traditional
 probabilistic interpretation
is not changed (notice that it
 practically did
not change during the last cca eighty years!). In the language of
textbooks one could speak just about the use of a non-unitary
generalizations $\Omega\sim \sqrt{\Theta}$
 of the most common Fourier transformations.
Still, the new physics behind the trick may be nontrivial (in
nuclear physics, for example, the mapping $\Omega$ (called Dyson's
\cite{Geyer}) was used to represent fermions as images of
 bosons).

Among the most innovative consequences of the  upgraded formulation
of quantum models one notices, first of all, the existence and
possibility of constructions of {a horizon} \cite{horizon}.
Formally, this notion coincides with the set {$\partial {\cal D}$}
of the Kato's \cite{Kato} exceptional points in the (real or
complex) manifold {${\cal D}$} of available free parameters (like
coupling strengths, etc).
The practical appeal of this notion may be based, e.g., on its
tunability \cite{tunable} and/or a new physics near instabilities
and quantum catastrophes \cite{bigbang}.

As another emergent concept one should list {fundamental length},
i.e., a quantity $\theta$ defined, in the simplified discrete
models, as the number of diagonals in the metric which is required
to possess
  {a band-matrix form},  {$\Theta^{(S)}_{mn} =0 \ {\rm for} \
   |m-n|>\theta$}.
In this context one might mention the first papers devoted to the
study of ${\cal PT}-$symmetric quantum graphs \cite{ptgraphs} in
which one might search for a connection between the fragile parts of
the spectrum and the topological characteristics of the underlying
graph structure.

Last but not least, it is necessary to emphasize the challenging
character of a generic scenario with more observables, each of which
may be responsible for its own part of the physical horizon,
 ``invisible" from the point of view of the other observables.
In other words, a lot of work is still to be done before one could
speak about a ``classification" of exceptional points (i.e., about a
a sort of ``quantum theory of catastrophes") -- the first attempts
in this direction only dealt with the hardly realistic, too
oversimplified and schematic quantum systems \cite{Chen}.

\section{
Coulomb potential $V(x_j)={\rm i}/x_j$}

The main weak point of the above-cited choice of the Coulomb
potential may be identified not only with its strict locality (i.e.,
with the {\em necessary} loss of the unitarity of the scattering,
cf. also the detailed study \cite{Jones} in this respect) but also
with the difficulties encountered during transition to {\em any}
model which would not be exactly solvable. For this reason, our
present use of the Runge-Kutta-inspired discretization will help
also in the case of the Coulomb potential.


As long as our present main ambition is the presentation of the
upgraded formalism, we shall try to simplify many inessential
mathematical aspects of our Coulomb/Kepler model. In parallel, we
shall also try to treat this potential as a special case of a
broader class of forces. For the sake of definitness and in a way
insspired by Ref.~\cite{annalso}, we shall pick up the class
$V(x)={\rm i}x^z$ with a real exponent which does not lie too far
from its Coulombic value of $z=-1$.

For this purpose we must replace, first of all, the typical
differential Schr\"{o}dinger equation
 \be
 -\frac{d^2}{dx^2}\,\psi_n(x)+
  V(x)\,\psi(x) =E_n\,\psi_n(x)\,
 \ \ \ \ \ \ \
 \psi(\pm \Lambda)=0\,,\ \ \ \ 1 \ll \Lambda \leq \infty
 \label{SEloc}
 \ee
by its discrete version (i.e., approximation or analogue)
 \be
 -\frac{\psi(x_{k-1})-2\,\psi(x_k)+\psi(x_{k+1})}{h^2}+V(x_k)\,
 \psi(x_k)
 =E\,\psi(x_k)\,.
 \label{SEdis}
 \ee
An equidistant grid of the Runge-Kutta points
 $
  x_k=-\Lambda+k\,h$ with $k = 0,  1, \ldots,
 N+1$ will be
used. In this sense, also the standard general double-integral inner
product will be replaced by the above-mentioned double sum, etc.

Naturally, the discretization recipe also involves the change of the
asymptotic
   boundary conditions, with
 $
  x_{N+1}=\Lambda$,  $h=2\Lambda/(N+1)$ and
  $
 \psi(x_{0})=\psi(x_{N+1})=0$.
In other words, the eigenvalue calculations become reduced to the
mere diagonalizations of the $N$ by $N$ matrix Hamiltonians
 \be
 {H}^{(N)}=
 \left (
 \begin{array}{ccccc}
 2+h^2V(x_{1})&-1&&&\\
 -1&2+h^2V(x_{2})&-1&&\\
  &-1& 2+h^2V(x_{3}) & \ddots&\\
  & &\ddots&\ddots&-1 \\
 &&&-1&2+h^2V(x_{N})
 \ea
 \right )\,.
 \label{dvanact}
 \ee
The insertion of any
 potential $V(x_j)={\rm i}x_j^z$ will
 lead to the eigenvalues $\varepsilon_j:=h^2E^{(N)}_j\in (0,4)$
 which must be computed numerically in general.

In some applications
 the transition to the continuous limit
$N=\infty$ is made or, at worst, {\em postponed} till the end of the
calculations. In the present methodical context we shall rather keep
the dimension $N$ constant and, in fact, not too large.


The insertion of formula $V(x_j)={\rm i}/x_j$ in Eq.~(\ref{dvanact})
with even $N=2K$ yields the sequence of the discrete ${\cal
PT}-$symmetric Coulomb Hamiltonians
 $ H^{(2K)}(a)=$
 \be
  \left[ \begin {array}{cccccccc}
 2-i\,a\,/(2K-1)&-1&0&\ldots&&&\ldots&0
 \\
 -1&\ddots&\ddots&\ddots&&&&\vdots
 \\
 0&\ddots &2-i\,a\,/3&-1&0&&&
 \\
 \noalign{\medskip}
 \vdots &\ddots&-1&2-ia&-1&0&&
 \\
 \noalign{\medskip}
 &&0&-1&2+ia&-1&\ddots&\vdots
 \\
 \noalign{\medskip}
 &&&0&-1&2+i\,a\,/3&\ddots &0
 \\
 \vdots &&&&\ddots&\ddots&\ddots&-1
 \\
 \noalign{\medskip}
 0&\ldots&&&\ldots&0&-1&2+i\,a\,/(2K-1)
 \end {array}
 \right]\,.
 \label{zzz}
 \ee
%
%
In its first nontrivial example let us set $N=4$,
 \be
 H^{(4)}(a,z)=
 \left[ \begin {array}{cccc}
 2-i\,a\,3^z&-1&0&0\\\noalign{\medskip}-1&2-ia&-1&0
 \\\noalign{\medskip}0&-1&2+ia&-1
 \\ \noalign{\medskip}0&0&-1&2+i\,a\,3^z
 \end {array}
 \right]\,,\ \ \ \ z=-1\,.
 \label{zz}
 \ee
We see that a natural generalization may be targeted not only at the
growing dimensions $N > 4$ but also towards the small deviations of
the exponent $z$ from its Coulombic value. Empirically, one can
verify that in both of these directions,
 the spectral loci (i.e., eigenvalues $\varepsilon^{(N,z)}(a)$)
 remain topologically the same. More precisely, at a fixed $N$, the
 topology of the
 Coulomb-potential pattern as sampled by Figs.~\ref{firm1} - \ref{firm6}
may be expected to survive all the  negative exponents $z$
\cite{annals}.

At $N=4$ the model is exactly solvable at the Coulombic exponent
$z=-1$. The secular equation
 $$
{{\it {E}}}^{4}-8\,{{\it {E}}}^{3}+ \left( 21+{\frac
{10}{9}}\,{a}^{2}
 \right) {{\it {E}}}^{2}+ \left( -{\frac {40}{9}}\,{a}^{2}-20 \right) {
\it {E}}+5+1/9\,{a}^{4}+5\,{a}^{2}=0
 $$
generates the closed-form spectrum
 $$
 \varepsilon(a)=2 \pm
 1/6\,\sqrt {54-20\,{a}^{2}\pm 2\,\sqrt {405-720\,{a}^{2}+64\,{a}^{4}}}
\,,
 $$
which is real iff $|a| \leq  3/4\,\sqrt {10-4\,\sqrt {5}} \approx
0.7706147226$.

At $N=6$ the model is still exactly solvable at $z=-1$, yielding the
secular equation
 $$
{{\it {E}}}^{6}-12\,{{\it {E}}}^{5}+ \left( 55+{\frac
{259}{225}}\,{a}^{ 2} \right) {{\it {E}}}^{4}+ \left( -120-{\frac
{2072}{225}}\,{a}^{2}
 \right) {{\it {E}}}^{3}+
 $$
 $$
 +\left( 126+{\frac {5894}{225}}\,{a}^{2}+{
\frac {7}{45}}\,{a}^{4} \right) {{\it {E}}}^{2}+ \left( -56-{\frac
{280 }{9}}\,{a}^{2}-{\frac {28}{45}}\,{a}^{4} \right) {\it
{E}}+
 $$
 $$
 +7+14\,{a}^{2 }+{\frac {7}{9}}\,{a}^{4}+{\frac
{1}{225}}\,{a}^{6}=0
 $$
which may be solved using Cardano formulae. Although the closed form
of the spectrum becomes extremely clumsy in this representation, it
decisively facilitates the graphical representation of the spectral
loci which all appear topologically equivalent to the vertical array
of circles. In particular, the survival of the exact solvability of
the problem enables us to conclude that the whole $N=6$ spectrum
remains real iff $|a| \leq 0.589586$.

These results indicate that the topological pattern remains generic
and $N-$independent. Such a conjecture is persuasively confirmed by
the larger$-N$ graphical samples which are presented in
Figs.~\ref{firm4} and \ref{firm6}.

\begin{figure}[h]                     
\begin{center}                         
\epsfig{file=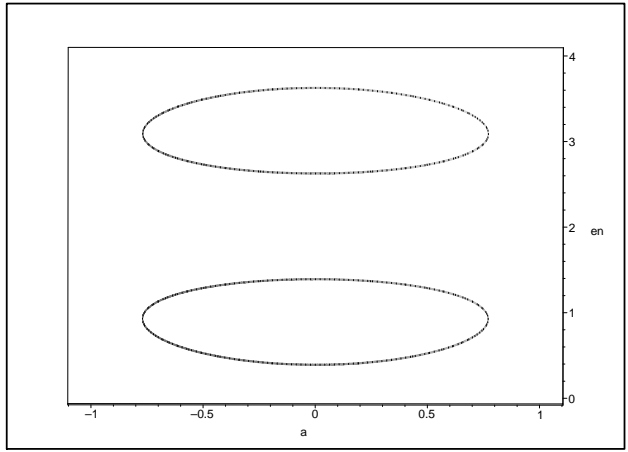,angle=270,width=0.6\textwidth}
\end{center}                         
\vspace{-2mm} \caption{The $a-$dependence of energies
$\varepsilon(a)$ at $N=4$.
 \label{firm1}}
\end{figure}

\subsection{Graphical methods}

Numerical evaluation of the spectra is sampled in Figs. \ref{firm1},
\ref{firm4} and
 \ref{firm6}. These
graphical constructions indicate that at any $N=2K$
 the  spectrum  is real iff $a \in \left
(-\alpha^{(2K)},\alpha^{(2K)}\right )$ and fully complex for $a
\notin \left (-\beta^{(2K)},\beta^{(2K)}\right )$. where
$\alpha^{(2K)} $ is a quickly decreasing function of $K$.

The latter observation may be interpreted in two ways. For the
finite lattices in which the numerical value of parameter $a$ is
fixed, the reality of the spectrum is, undoubtedly, fragile. In the
alternative approach in which our model serves just as a simulation
of the (NB: exactly solvable!) differential-equation system, the
definition of parameter $a$ is prescribed by the Runge-Kutta recipe
(see above). For this reason, its numerical value decreases, with
$N$, much more quickly than $\alpha^{(N)}$. This implies that in the
latter setting the reality of the spectrum may be declared robust
and guaranteed.

\begin{figure}[h]                     
\begin{center}                         
\epsfig{file=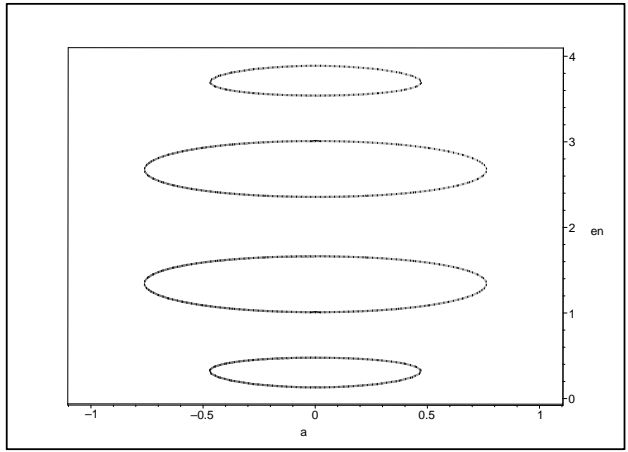,angle=270,width=0.6\textwidth}
\end{center}                         
\vspace{-2mm} \caption{The $a-$dependence of energies
$\varepsilon(a)$ at $N=8$.
 \label{firm4}}
\end{figure}

Marginally, we may add that in the former scenario using small and
fixed $N$, the loss of the reality of the spectrum is caused by the
confluence of the ground state with the first excited state and by
their subsequent complexification. Due to the up-down symmetry of
the spectrum, this instability is paralleled by the upper two states
of course.

\begin{figure}[h]                     
\begin{center}                         
\epsfig{file=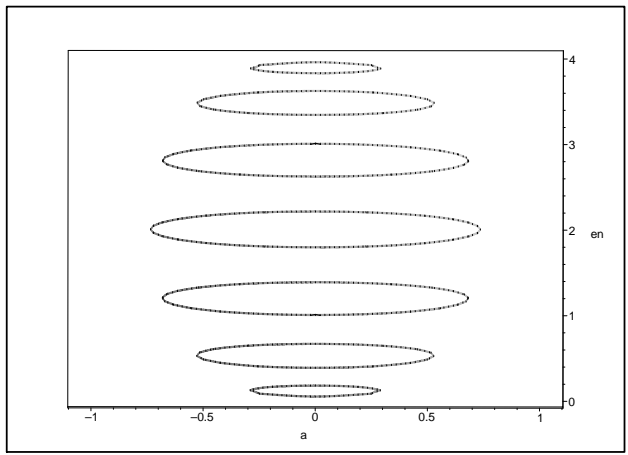,angle=270,width=0.6\textwidth}
\end{center}                         
\vspace{-2mm} \caption{The $a-$dependence of energies
$\varepsilon(a)$ at $N=14$.
 \label{firm6}}
\end{figure}

\section{Hermitizations \label{metrics}}

\subsection{The  $N=2$ metrics \label{nje2}}

The condition of hidden Hermiticity of {\em any} Hamiltonian with
real spectrum is often called Dieudonn\'e equation \cite{SIGMA},
 \be
   H^{(N)}=\left
 [H^{(N)}\right ]^\ddagger:=
 \Theta^{-1}\left [H^{(N)}\right ]^\dagger\Theta\,.
 \label{dieudonne}
 \ee
Its structure is best illustrated at $N=2$ where the metrics form
just the two-parametric family
 \be
 \Theta\left [H^{(N)}(a) \right ]=\Theta\left [H^{(N)}(a) \right
 ]_{(k,m)}=\left[ \begin {array}{cc} k&km-ika
 \\\noalign{\medskip}km+ika&k
 \end{array} \right]\,,\ \ k,m \in \mathbb{R}\,.
 \label{pse2}
 \ee
With the  condition $a \in (-1,1)$ of the reality of energies and
with the metric-positivity constraint
 \be
 \theta=\theta_{\pm}=k\pm \sqrt {{k^2m}^{2}+{k}^{2}{a}^{2}}>0\,
 \label{positivitym}
 \ee
we may conclude that
 $k$ must be positive  and larger than the square root.
We may reparametrize  $a=\cos \beta\,\sin \gamma$,
 $m=\cos \beta\,\cos \gamma$,
$\beta\in (0,\pi)$ and
 $\gamma\in (0,\pi)$ and $-1<\cos \beta<1$
and get the final result
 \be
 \Theta=
 \Theta\left \{H^{(2)}[a(\beta,\gamma)]
 \right \}_{[k,m(\beta,\gamma)]}=k\cdot\left[
 \begin {array}{cc} 1&e^{-{\rm i}\gamma}\cos \beta
 \\\noalign{\medskip}e^{{\rm i}\gamma}\cos \beta&1
 \end{array} \right]\,.
 \label{metrika2}
 \ee
In a search for the {\em other} eligible observables with
crypto-Hermiticity property
 \be
 \Lambda_{(\beta,\gamma)}^\dagger\,\Theta_{[k,m(\beta,\gamma)]}
 =\Theta_{[k,m(\beta,\gamma)]}\,\Lambda_{(\beta,\gamma)}\,.
 \label{diela}
 \ee
the use of the ansatz
 \be
 \Lambda=\left[
 \begin {array}{cc}
 G+{\rm i}g&B+{\rm i}b
 \\\noalign{\medskip}C+{\rm i}c&D+{\rm i}d
 \end{array} \right]\,.
 \label{leila}
 \ee
leads to the
 four real constraints imposed upon eight free parameters.
The
 family of
observables is four-parametric, therefore. Three constraints define
$B$, $C$ and  $G-D$. The
 remaining one
relates the sums $c_\Sigma=b+c$ and $g_\Sigma=g+d$ and leads to the
 unique solution $g_\Sigma=0$.
We may conclude that from the input $m=m(\beta,\gamma)$ and
$a=a(\beta,\gamma)$ one gets the class of admissible observables
 \be
  \Lambda=\Lambda(D,b,c,g)=\frac{1}{a}\cdot \left[ \begin {array}{cc}
   D\,a- b-c+i\,g\,{a}\,,
   &
   g-b\, m+i \,b\,a
 \\\noalign{\medskip}
 {g+c\,m}+i\,c\,a\,,
 &
 D\,a-i\,g\,a
 \end {array} \right]\,.
 \label{geleila}
 \ee
In particular, the initial Hamiltonian is reobtained at $D=2$,
$G=2$, $b=c=0$, $B=C=-1$ and $g= -a\ ( =-d)$.

In the literature the concept of charge is considered particularly
useful \cite{Carl}. Its essence lies, in the present model, in an
{\em additional} auxiliary assumption
 \be
 \left [H\right ]^\dagger{\cal P}={\cal P}\,H
 \,
 \label{pseudodieudonne}
 \ee
where ${\cal P}$ is the operator of parity. Under this assumption a
{\em unique} metric is sought such that a very specific metric
 called ${\cal CPT}$ metric
is prescribed by formula
 $\Theta^{({\cal CPT})}={\cal
CP}$ where ${\cal C}$ is called ``charge".

Thus, at a given $N$ we may define the parity ${\cal P}={\cal
P}^{(N)}$ which contains units along the secondary diagonal, i.e.,
 $ {\cal P}^{(N)}_{m,n}= 1$ iff $m+n=N+1$ while ${\cal
P}^{(N)}_{m,n}= 0$ otherwise. Incidentally one may  note that
 ${\cal P}^{(2)}$ =
limit of $\Theta\left [H^{(2)}(a) \right ]$ such that $k^{({\cal
P})} \to 0$, $m^{({\cal P})} \to \infty$, $k^{({\cal P})}m^{({\cal
P})} \to 1$.

The key merit of the use of charge is that its use makes the metric
unique,
 \be
 {\cal C}^{({\cal CPT})}=k \cdot \left[
 \begin {array}{cc}
 - {\rm i} a& 1
 \\\noalign{\medskip}1& {\rm i} a
 \end{array} \right]\,.
 \label{uleila}
 \ee
Moreover, it also represents one of the special cases of  observable
$\Lambda$ using $D=b=c=0$ and $g = - \cos \beta/\sin \beta=-
\sqrt{k^2-1}$. Indeed, from $\Theta^{({\cal CPT})}={\cal CP}$ we
have
 \be
 {\cal C}=\Theta\left [H^{(2)}(a) \right ]_{(k,m)}{\cal P}^{(2)}=\left[
 \begin {array}{cc}
 u&v
 \\\noalign{\medskip}y&z
 \end{array} \right]\,
 \label{uleila}
 \ee
yielding $v=y=k$ and $z=u^*=ke^{i\gamma}\cos \beta$. Then, condition
${\cal C}^2=I$ requires that $\gamma=\gamma^{({\cal CPT})}= \pi/2$
(i.e.,  $a=\cos \beta$). Thus, we have $\beta=\beta^{({\cal CPT})}$
such that $\sin \beta^{({\cal CPT})}=1/k$. This proves the above
statement.

\subsection{The  $N=4$ metrics \label{nje4}}

With the natural ansatz for $\Theta\left [H^{(4)}(a,z) \right
]_{(k,m,r,h)}=$
 \be
 =\left[ \begin {array}{cccc}
 k&m-ikw&W^*&Z^*
  \\\noalign{\medskip}m+ikw&r&h-i
 \left( kw+ra \right)&W^*
 \\\noalign{\medskip}W &h+i
 \left( kw+ra \right) &r&m-ikw
 \\\noalign{\medskip} Z&W &m+ikw&k
 \end {array} \right]
 \label{gene4}
 \ee
where  $w=w(z,a) = {3}^{z}{a}\ $  and
 $$
 W=W(k,m,r)=-{w}^{2}k+r-k- kwa+i \left( wm+ma \right)\,,
 $$
 $$
 Z=Z(k,m,r,h)=m{a}^{2}-{w}^{2}m-m+h-i \left( kw-ka-kw{a}^{2}-rw+ {w}^{3}k
 \right)\,,$$
the problem of the determination of the domain of positivity of the
metric starts to be merely tractable graphically.

For the numerous practical purposes the metric is sought in a
special form. One of the phenomenologically inspired options is the
choice of the matrix  with  units along its main diagonal, $k=r=1$.
In addition, let us select $m=h=0$ and compute $W(1,0,1)=-{w}\left(
w+a \right)$, $Z(1,0,1,0)=i \left( a+w{a}^{2}- {w}^{3} \right)$.
Such a restricet construction leads to the following four
closed-form eigenvalues of the metric $\Theta \left[ H^{(4)}(a,z)
\right ]_{(1,0,1,0)}$, viz,
 $$\theta_+^\pm
 =1+  \frac{1}{2}\,\left (w-{a}^{2}w+{w}^{3}\right )
 \pm \frac{1}{2}\,\sqrt{\triangle^+}
 $$
 $$\theta_-^\pm
 =1-  \frac{1}{2}\,\left (w-{a}^{2}w+{w}^{3}\right )
  \pm \frac{1}{2}\,\sqrt{\triangle^-}
 $$
 $$
 \triangle^\pm=
 {w}^{6}+ \left( 2-2\,{a}^{2} \right) {w}^{4}+ \left( \pm 8+4\,a \right) {w
}^{3}+ \left(5 \pm 8\,a+ 6\,{a}^{2}+{a}^{4} \right) {w}^{2}+
 $$
 $$
 +\left( 4\,a+4 \,{a}^{3} \right) w+4\,{a}^{2}\,.
 \ \ \ \ \ \ \ \ \ \ \ \ \ \
 \ \ \ \ \ \ \ \ \ \ \ \ \ \
 \ \ \ \ \ \ \ \ \ \ \ \ \ \
 $$
Some details and numerical results of its analysis may be found
elsewhere \cite{annals}.

\section{Discussion}

In a climax of our present discussion of the discrete ${\cal
PT}-$symmetric Coulomb problem characterized by the purely imaginary
coupling constant, let us now summarize the overall method via the
following scheme
 \ben
  \ba
    \begin{array}{|c|}
 \hline
 \ {\rm textbook \  level\ quantum\ theory:}\ \\
 {\rm {} prohibitively\ complicated}\ {\rm \ Hamiltonian}\ \mathfrak{h}\\
 {\rm generating \ unitary\  time\ evolution}\ \\
    \ \ \ \ {\rm \fbox{P:} \ {{} physics\  = \  trivial} } \ \\
  \ \ \ \ \fbox{\rm  calculations\ =\  practically\ impossible\ } \ \\
 \hline
 \ea
 \\
 \stackrel{{}  simplif\/ication}{}
 \ \ \ \
  \swarrow\ \  \  \ \ \ \ \ \
 \ \ \ \ \ \ \ \
  \ \  \  \ \ \ \ \ \
 \ \ \ \ \  \searrow \nwarrow\ \ \
 \stackrel{{}  unitary\ equivalence}{}\\
 \begin{array}{|c|}
 \hline%
 \ {\rm  \ state}\ \psi \ {\rm is\ represented}\\
 \
  {\rm in\ the} \ \fbox{{} false\  {\rm Hilbert\ space}}\ \\
 \
  {\rm   \fbox{F:} \ calculations\ = \ {{}  feasible}\ }\    \\
 \
  {\rm  physical\ meaning \ = \ lost}\  \\
  \ \ \ \ H = {\rm non-Hermitian\ } \ \\
  \hline
 \ea
 \stackrel{ {{}  hermitization}  }{ \longrightarrow }
 \begin{array}{|c|}
 \hline
 \ \fbox{\rm amended \  inner\ product}\  \\
 \
  {\rm {} standardized} {\rm \ representation } \
 \\
  \ \
  {\rm  \fbox{S:} \ picture\ }
  =
   \  {\rm {} synthesis\ } \  \\ %
  \ \
  {\rm   physics\ = \ reinstalled}\   \\
  \ \ \ \ H = {\rm Hermitian\ } \ \\
  %
 \hline
 \ea
\\
\\
\ea
 \een
which characterizes the ``three-Hilbert-space" pattern of
quantization as described in Ref.~\cite{SIGMA} as a recipe in which
the usual Schr\"{o}dinger equation
 \be
 H\,|\psi_n\kt = E_n\,|\psi_n\kt\,
 \ee
finds the standard probabilistic interpretation even if the
Hamiltonian matrix (with real spectrum) proves manifestly
non-Hermitian.

The specific feature of non-Hermitian matrices $H$ may be seen in
their ability of having the reality of their spectra {\em
controlled} by a parameter (for this purpose we used $a$ in our
present models). In other words, one can simulate the abrupt loss of
the stability of the time evolution of the system by a mere smooth
change of this parameter. In other words, we may speak about
 a non-empty (quasi-)Hermiticity domain of parameters,
 with the {\em qualitative} changes of physics at its boundary, and with
 a guaranteed reality of the spectrum in its interior.

In our present paper we emphasized that another important aspect of
physics with real spectra but non-Hermitian matrices of observables
lies in the necessity of a fine-tuning mediated by the Dieudonne
equation.

Typically, a {\em given} Hamiltonian $H$ must be assigned a
Hermitizing metric $\Theta$. As long as we merely considered
$N<\infty$, we could avoid any difficulties by simply solving the
second, conjugate Schr\"{o}dinger equation
 \be
  \bbr \psi_m|\,H=F_m\,\bbr \psi_m|\,
 \ee
which may be also written in the form
 $$
 H^\dagger\,|\psi_m\kkt = F_m^*\,|\psi_m\kkt\,
 $$
This enabled us to work with the solutions as forming a bicomplete
and  biorthogonal basis,
 \be
 I = \sum_{n=0}^{N-1}\,|\psi_n\kt\,
 \frac{1}{\bbr \psi_n|\psi_n\kt}\,\bbr \psi_n|\,,
 \ \ \ \ \
 \bbr \psi_m|\psi_n\kt = \delta_{m,n}\,\bbr \psi_n|\psi_n\kt\,.
 \ee
The main benefit may be then found in the  {closed formula}
$$
 \Theta= \sum_{n=0}^{N-1}\,|\psi_n\kkt\,
 |\kappa_n|^2\,\bbr \psi_n|\,
 $$
which defines {all of the eligible} metrics.
%
%
%
This, in its turn, specifies all the dynamics given by the operator
doublet $(H({\lambda}),\Theta({\kappa}))$.

\subsection*{Acknowledgment}

Work supported by the GA\v{C}R grant Nr. P203/11/1433.


\begin{thebibliography}{00}

\bibitem{Hook}
Hook D (2012) The PT Symmeter. http://ptsymmetry.net. Cited 30 Mar
2012

\bibitem{SIGMA}
Znojil M (2009) Three-Hilbert-space formulation of Quantum
Mechanics. SIGMA 5: 001 (19 pp), arXiv:0901.0700

\bibitem{Carl}
Bender C M (2007)
 Making sense of non-Hermitian Hamiltonians.
Rep. Prog. Phys. 70: 947-1018

\bibitem{BG}
Buslaev V, Grechi V (1993)
 Equivalence of unstable
 anharmonic oscillators and double wells.
J. Phys. A: Math. Gen. 26: 5541–5549

\bibitem{Langer}
Langer H, Tretter Ch (2004) A Krein space approach to PT symmetry.
Czech. J. Phys. 54: 1113-1120

\bibitem{Str}
Streater R F, Wightman  A S (1964) PCT, spin and statistics, and all
that. Benjamin/Cummings, London. ISBN 0-691-07062-8.

\bibitem{BB}
Bender C M, Boettcher S (1998)
 Real spectra in non-Hermitian
 Hamiltonians having PT symmetry.
Phys. Rev. Lett. 80: 5243-5246

\bibitem{ptcoul}
Znojil M, L\'evai G (2000) The Coulomb - harmonic-oscillator
correspondence in PT symmetric quantum mechanics. Phys. Lett. A 271:
327-333

\bibitem{geza}
Znojil M, Siegl P, L\'evai G (2009) Asymptotically vanishing
PT-symmetric potentials and negative-mass Schroedinger equations.
 Phys. Lett. A  373:  1921–1924

\bibitem{siegl}
L\'evai G, Siegl P, Znojil M (2009)
 Scattering in the PT-symmetric Coulomb potential.
J. Phys. A: Math. Theor. 42: 295201 (9pp)

\bibitem{annals}
Znojil M (2012) N-site-lattice analogues of $V(x)=i x^3$. Ann. Phys.
(NY) 327: 893-913

\bibitem{tobog}
Znojil M (2011), Planarizable supersymmetric quantum toboggans.
SIGMA 7: 018 (24 pp), doi:10.3842/SIGMA.2011.018, arXiv:1102.5162

\bibitem{cubic}
Mostafazadeh A (2006) Metric operator in pseudo-Hermitian quantum
mechanics and the imaginary cubic oscillator. J. Phys. A: Math. Gen.
39: 10171-10188

\bibitem{ali}
Mostafazadeh A (2010)
 Pseudo Hermitian representation of quantum
 mechanics.
Int. J. Geom. Meth. Mod. Phys. 7: 1191-1306

\bibitem{Bohm}
Bohm A R, Gadewlla M, Kielanowski P (2011) Time Asymmetric Quantum
Mechanics. SIGMA 7: 086 (13 pp)

\bibitem{Hoo}
Hilgevoord J (2001) Time in Quantum Mechanics. Am. J. Phys. 70: 301
- 306

\bibitem{DDT}
Dorey P, Dunning C, Tateo R (2001)
 Spectral equivalences, Bethe
Ansatz equations, and reality properties in PT-symmetric quantum
mechanics. J. Phys. A: Math. Gen. 34: 5679-5704

\bibitem{Geyer}
Scholtz F G, Geyer H B, Hahne F J W (1992) Quasi-Hermitian Operators
in Quantum Mechanics and the Variational Principle. Ann. Phys. (NY)
213: 74-101

\bibitem{Schomerus}
Schomerus H (2011) Universal routes to spontaneous PT-symmetry
breaking in non-hermitian quantum systems. Phys. Rev. A 83:
030101(R)

\bibitem{Longhi}
R\"{u}ter C E, Makris R, El-Ganainy K G, et al (2010)
Observation of parity-time symmetry in optics. Nat. Phys. 6: 192-195

\bibitem{Scott}
V. Jakubsk\'{y} V (2007) Thermodynamics of pseudo-Hermitian systems
in equilibrium. Mod. Phys. Lett. A 22: 1075-1084

Joglekar Y N, Karr W A (2011) Phys. Rev. E 83:  031122

\bibitem{scatt}
Znojil M (2008) Scattering theory with localized non-Hermiticities.
Phys. Rev. D 78: 025026,
 doi: 10.1103/PhysRevD.78.025026

\bibitem{Jones}
Jones  H F (2007) Scattering from localized non-Hermitian
potentials. Phys. Rev. D 76:  125003 (5pp)

\bibitem{fund}
Znojil M (2009) Fundamental length in quantum theories with
PT-symmetric Hamiltonians. Phys. Rev. D 80: 045022

\bibitem{horizon}
Znojil M (2008) Horizons of stability.
 J. Phys. A: Math. Theor. 41:  244027

\bibitem{Kato}
Kato T (1966) Perturbation theory for linear operators. Springer,
Berlin

\bibitem{tunable}
Znojil M (2007) A return to observability near
 exceptional points in a schematic PT-symmetric model.
Phys. Lett. B 647: 225-230

\bibitem{bigbang}
Znojil M (2012) J Phys: Conf. Ser. 343: 012136 (20 pp)

\bibitem{ptgraphs}
Znojil M (2009) Fundamental length in quantum theories with
PT-symmetric Hamiltonians II: The case of quantum graphs. Phys. Rev.
D. 80: 105004 (20 pp)

\bibitem{Chen}
Chen J-H, Pelantov\'{a} E, Znojil M (2008) Classification of the
conditionally observable spectra exhibiting central symmetry. Phys.
Lett. A 372: 1986-1989


\end{thebibliography}
\end{document}